\documentclass[pra,floatfix,twocolumn,tightenlines,nofootinbib,superscriptaddress]{revtex4-1}

\usepackage[utf8]{inputenc} 
\usepackage{amsmath,amsthm,amssymb}
\usepackage{graphicx}
\usepackage{times}
\usepackage{hyperref}
\hypersetup{
colorlinks=true,
linkcolor=red,
citecolor=red,}
\usepackage{textcomp}
\usepackage{xcolor}
\usepackage{enumerate}
\usepackage{multirow}
\usepackage{tabularx}
\usepackage{dcolumn}
\usepackage[mathscr]{euscript}
\usepackage{mathtools}
\usepackage{textcomp}
\usepackage[separate-uncertainty=true]{siunitx}
\usepackage{soul}
\usepackage{comment}
\usepackage{dsfont}
\usepackage{amsfonts}
\newcommand\s[1]{_{\rm #1}}
\newcommand\sn[2]{_{{\rm #1}{|}#2}}
\newcommand\su[1]{^{(\rm #1)}}
\newcommand{\ud}{^{\rm d}}

\newcommand{\bra}[1] {\langle #1 |}
\newcommand{\ket}[1] {| #1 \rangle}
\newcommand{\braket}[2] {\langle #1 | #2 \rangle}
\newcommand{\ketbra}[1]{ | #1 \rangle\!\langle #1 |}
\newcommand{\sketbra}[2]{ | #1 \rangle_{\rm #2}\langle #1 |}

\newcommand{\Tr} {\operatorname{Tr}}

\newcommand{\one}{\leavevmode\hbox{\small1\normalsize\kern-.33em1}}
\newcommand{\sandwich}[3]{\mbox{$ \langle #1 | #2 | #3 \rangle $}}

\newcommand{\id}{\mathsf{I}}
\newcommand{\Ident} {\mathds 1}
\newcommand{\M}{{\cal M}}

\newcommand{\St}{{\cal S}}

\newcommand{\B}{{\cal B}}
\newcommand{\C}{{\cal C}}

\newcommand{\MM}{\mathsf{M}}

\newcommand{\trs}{^\mathsf{T}}

\begin{document}
\title{Incoherent witnessing of quantum coherence}
 \author{Sahar Basiri-Esfahani}
 \email{Electronic address: sahar.basiriesfahani@swansea.ac.uk}
 \affiliation{Department of Physics, Swansea University, Singleton Park, Swansea SA2 8PP, United Kingdom}

 \author{Farid Shahandeh}
 \email{Electronic address: shahandeh.f@gmail.com}
 \affiliation{Department of Physics, Swansea University, Singleton Park, Swansea SA2 8PP, United Kingdom}
\date{\today}

\begin{abstract}

Theoretical and experimental studies have suggested the relevance of quantum coherence to the performance of photovoltaic and light-harvesting complex molecular systems.
However, there are ambiguities regarding the validity of statements we can make about the coherence in such systems.
Here we analyze the general procedure for coherence detection in quantum systems and
show the counterintuitive phenomenon of detecting a quantum system's initial coherence when both the input and output probe states are completely incoherent. 
Our analysis yields the necessary and sufficient conditions for valid claims regarding the coherence of directly inaccessible systems. 
We further provide a proof-of-principle protocol that uses entangled probes to detect quantum coherence satisfying these conditions, and discuss its potency for detecting coherence.
\end{abstract}

\maketitle

Quantum coherence---that quantum systems occupy multiple states simultaneously and hence exhibit interference---is the distinctive feature of quantum systems compared to classical ones.
In recent years, quantum coherence has become a critical element of developing quantum technologies that aim at improving over classical approaches~\cite{Engel2013Science,Scholes2017Nature,scully2010PRL,kimble2008Nature,Streltsov2017}.
Of particular importance is the potential role of quantum coherence in the light-harvesting efficiency of biochemical processes, e.g., photosynthesis~\cite{Scholes2011Nature,Kassal2016JPCL,Scholes2010JPCL,Scholes2015ARPC}, as well as enhancing the performance of molecular systems such as organic solar cells~\cite{Kassal2019PRA,Kassal2020JPCL,GaugerPhotocell2016PRL,KassalGauger2021JPCL,ChinPhotocell2013PRL}.
This has stimulated studies of the effects of quantum coherence in such systems, at the heart of which lies schemes to certify the presence of quantum coherent mechanisms in molecular systems~\cite{Fleming2007Nature,Scholes2010Nature,SMukamel2011PNAS,SMukamel2010JPC,Engel2010PNAS}.

\begin{figure}[t!]
\centering
\includegraphics[width=0.7\linewidth]{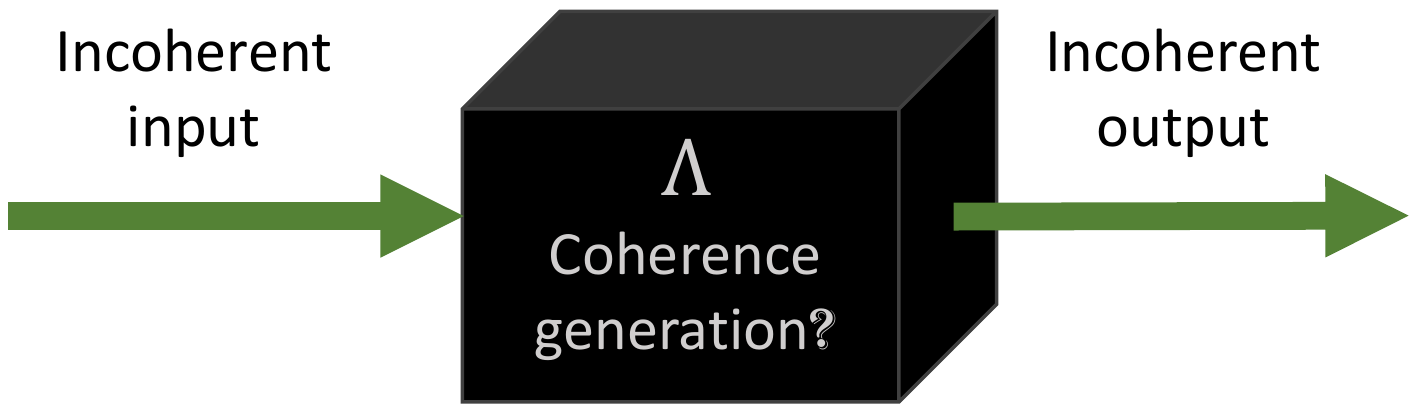}
\caption{A coherence-generating quantum channel with incoherent input and output.
In such circumstances, the quantum coherence generated in a directly inaccessible system seems undetectable.}
\label{fig1}
\end{figure}

The main tool to examine quantum coherence in complex systems including photosynthetic complexes is the ultrafast spectroscopy~\cite{Fleming2007Nature,Scholes2010Nature,SMukamel2011PNAS,SMukamel2010JPC,Engel2010PNAS}.
Spectroscopic observations, however, have led to debates~\cite{kassal2013JPCL,Engel2019Nat.Rev.Chem,PBrumer2012PNAS} mainly because the dynamics of systems in nature, as opposed to the spectroscopic techniques, is initiated by incoherent inputs such as sunlight~\cite{PBrumer2012PNAS,manvcal2010,PBrumer1991}.
These arguments indicate the need for further investigating the existence of quantum coherence in systems operating under ambient conditions and proposals of new protocols to detect quantum coherence using incoherent light sources rather than coherent lasers~\cite{FrankSchlawin2020PRL}.
Furthermore, it is intuitive to assume that when quantum channels suffer from significant decohereing noise the coherence of the system is untraceable. 
We approach these arguments from a quantum informational perspective and pose the following question (Fig.~\ref{fig1}):
Is it possible to make deductions about the coherence properties of a system when both the input and output probe states are incoherent?

Here, we answer to the above question in the affirmative.
We show via a counter-example that quantum coherence of a system can be observed even if both input and output to the process are fully incoherent. We also provide rigorous necessary and sufficient conditions for this observation to be possible.
We propose a proof-of-principle protocol to detect the existence of quantum coherence within a system using incoherent light sources and discuss the physics behind this counter-intuitive phenomenon.
%%%%%
%%%%%
Let us begin with the description of quantum coherence from the perspective of quantum information science~\cite{Streltsov2017}.
In this picture, quantum coherence comprises of the following ingredients.
First, given a system described in a finite-dimensional Hilbert space, an orthonormal basis
$\B\s{inc}{=}\{|i\rangle\}$ is defined as the {\it incoherent basis}.
These vectors represent the basis with respect to which we would like to consider the quantum coherence properties of the system.
Since basis vectors $\ket{i}$ are also pure quantum states, they define all {\it incoherent} quantum states as probabilistic mixtures of the incoherent basis states.
In other words, every density operator $\varrho\s{inc}$ that can be written as $\varrho\s{inc} {=} \sum_i p_i \ketbra{i}$, for $\ket{i}{\in} \B\s{inc}$, $p_i{\geq} 0$, and $\sum_i p_i {=} 1$, is an incoherent state.
The terminology is indeed justified by noting that every $\varrho\s{inc}$ is diagonal in the basis $\B\s{inc}$ with no off-diagonal matrix elements. 
We denote the collection of incoherent density operators by $\St\s{inc}$.

The second ingredient of the theory is the set of {\it incoherent channels} denoted by $\C\s{inc}$.
A quantum channel is a transformation that converts any input density operator to an output density operator.
Most generally, a channel is called incoherent if upon receiving an incoherent input quantum state outputs an incoherent quantum state.
Several classes of such transformations have been studied within the quantum information literature so far~\cite{Chitambar2016-2,Streltsov2017}.

In order to better understand how these channels look like, it is useful to introduce a particular transformation called the {\it fully dephasing} channel $\Delta$ that acts on any quantum state as $\Delta[\varrho]:=\sum_i \sandwich{i}{\varrho}{i} \ketbra{i}=\sum_i \varrho_i \ketbra{i}$. 
Here, we have introduced the shorthand $\varrho_i{:=}\sandwich{i}{\varrho}{i}$ for the diagonal matrix elements of the density operator $\varrho$ in the incoherent basis $\B\s{inc}$, or simply the {\it population} of each incoherent state of the system.
It is thus evident that the action of the fully dephasing channel is to completely suppress the coherence of the input quantum state.
Incoherent channels can now be defined as all channels $\Lambda$ that satisfy~\cite{Chitambar2016,Chitambar2016-2}
\begin{equation}\label{eq:MIO}
    \Delta\circ\Lambda[\varrho\s{inc}]=\Lambda[\varrho\s{inc}]
\end{equation}
for all incoherent input quantum states $\varrho$.
Here, $\circ$ denotes the {\it composition} of quantum channels, meaning that each channel is consecutively applied to the output of the previous channel to its right.
Equation~\eqref{eq:MIO} has a very intuitive physical interpretation: a channel is incoherent if and only if the further application of a fully dephasing channel on its output is redundant for every incoherent input.
Equation~\eqref{eq:MIO} can also be rewritten in terms of {\it all} input quantum states as
\begin{equation}\label{eq:DeltaMIO}
    \Delta\circ\Lambda\circ\Delta[\varrho]=\Lambda\circ\Delta[\varrho],
\end{equation}
wherein a fully dephasing channel has been used to transform the arbitrary input quantum state into an incoherent one.

The third and final ingredient of the theory of quantum coherence are {\it incoherent measurements}.
It is known that every quantum measurement is described by a set of {\it effects} $\MM{=}\{M_k\}$.
An effect $M_k$ is positive, i.e., it is Hermitian with nonnegative eigenvalues, and corresponds to the outcome $k$ of the measurement $\MM$.
The collection $\MM$ such that $\sum_k M_k{=}\Ident$ is called a positive operator-valued measure (POVM)~\cite{NielsenBook}.
Furthermore, according to the Born rule, given the quantum state $\varrho$ the probability of outcome $k$ in measurement $\MM$ is given by $p(k|\varrho,\MM)=\Tr M_k\varrho$.

Now, a measurement $\MM\s{inc}$ is called incoherent if the matrix representation of each of its effects is diagonal with respect to the incoherent basis $\B\s{inc}$~\cite{Streltsov2017}.
The simplest example of an incoherent measurement is indeed the projective measurement onto the incoherent basis, i.e. $\MM\s{inc}{=}\{P_i{=}\ketbra{i}\}{:=}\Pi\s{inc}$.
We denote the collection of all incoherent measurements by $\M\s{inc}$.

{\it Channels and the probe-system interaction}.---
The concept of a channel is very versatile.
Consider the scenario in which a probe interacts with a system and is then measured.
It is implicit that the system's degree of freedom of interest cannot directly be accessed without the mediation of the probe.
As a result, all we can speak of is the initial and the final quantum state of the probe.
In other words, the system---including its initial and final quantum states---and its interaction with the probe are subsumed by the quantum channel that transforms the probe.
 
A generic channel $\Lambda$ transforming an input state to the output can be written as~\cite{NielsenBook}
\begin{equation}\label{eq:channel}
    \varrho\su{out}=\Lambda [\varrho\su{in}]=\sum_i K_i\varrho\su{in} K_i^\dag.
\end{equation}
Here, the operators $ K_i$ are called Kraus operators of the channel and satisfy $\sum_i  K_i^\dag  K_i{=}\Ident$.
This way, it is guaranteed that the output is a valid normalized density operator; see Supporting Information-1 (SI-1).
The simplest example of a quantum channel is indeed a unitary one that acts on the input as
\begin{equation}\label{eq:unitarychannel}
    \varrho\su{out}=\Upsilon [\varrho\su{in}]= U\varrho\su{in} U^\dag.
\end{equation}

An important question following Eq.~\eqref{eq:channel} is that what happens to the unitarity and reversibility of quantum mechanics.
A fundamental theorem of quantum mechanics implies that every generic quantum channel as in Eq.~\eqref{eq:channel} can be {\it purified} to a unitary channel acting on the input state and some {\it inaccessible} auxiliary system~\cite{NielsenBook}, that is,
\begin{equation}\label{eq:unitarypurif}
    \varrho\su{out}=\Lambda [\varrho\su{in}]=\Tr\s{a}\Upsilon[\varrho\su{in}\otimes \pi\s{a}\su{in}].
\end{equation}
In Eq.~\eqref{eq:unitarypurif}, $\pi\s{a}$ is a suitably chosen state of the ancillary subsystem ${\rm a}$, $\Upsilon$ is a suitably chosen interaction unitary channel acting on the joint input-ancilla compound, and $\Tr\s{a}$ is the partial trace taken over the ancilla.
We note that the partial trace represents the inaccessibility of the auxilliary subsystem to the experimenter.

Equation~\eqref{eq:unitarypurif} can be brought to the context of probe-system interaction by interpreting the input subsystem as the {\it probe} (from now on denoted as $\varrho\s{p}$) and the ancilla subsystem as the {\it system of interest} (from now on denoted as $\pi\s{s}$).
Neglecting the interactions with the environment for the moment, $\Upsilon$ can be interpreted as the unitary interaction between the probe and the system, denoted by $\Upsilon\s{ps}$.
The inaccessibility of the ancilla thus translates to the fact that direct measurements on the system of interest are out of rich.
Eq.~\eqref{eq:unitarypurif} can thus be rewritten as
\begin{equation}\label{eq:UP}
    \varrho\s{p}\su{out}=\Lambda [\varrho\s{p}\su{in}]=\Tr\s{s}\Upsilon\s{ps}[\varrho\s{p}\su{in}\otimes \pi\s{s}\su{in}],
\end{equation}
that allows us to connect the properties of the inaccessible system to the properties of the channel acting on the probe alone.

In the present work we are interested in the {\it initial} coherence properties of the system $\pi\s{s}\su{in}$ by merely observing
the output probe state $\varrho\s{p}\su{out}$.
It is evident from Eq.~\eqref{eq:UP} that there are three elements that can cause observable coherent effects in the probe, namely, the initial system state $\pi\s{s}\su{in}$, the initial probe state $\varrho\s{p}\su{in}$, and the probe-system interaction $\Upsilon\s{ps}$.
Thus, to make valid statements about the coherence of $\pi\s{s}\su{in}$, we must make sure the latter two potential causes are not in effect by making two assumptions:
\begin{enumerate}[(i)]
    \item The input state of the probe is incoherent;
    \item The channel $\Lambda$ is incoherent for all incoherent input states of the system and probe.
\end{enumerate}

\noindent We note that assumption (i) necessarily prevents objections of the kind associated with spectroscopic techniques~\cite{kassal2013JPCL,Engel2019Nat.Rev.Chem,PBrumer2012PNAS,manvcal2010,PBrumer1991}.
Now, let us posit the appropriate incoherent bases for the probe and the system to be $\B\s{inc;p}$ and $\B\s{inc;s}$, respectively.
Assumption (ii) can then be expressed as
\begin{equation}\label{eq:unitarycond}
\sum_\phi {\s{ps}\!}\bra{k,\phi} U\s{ps}\sketbra{i',j'}{ps} U\s{ps}^\dag \ket{l,\phi}\s{ps} = 0,
\end{equation}
for all $\ket{i'}\s{p},\ket{k}\s{p},\ket{l}\s{p}\in\B\s{inc;p}$ with $k\neq l$ and all $\ket{j'}\s{s}\in\B\s{inc;s}$.
Furthermore, $\{\ket{\phi}\s{s}\}$ is an arbitrary basis for the Hilbert space of the system.
Equation~\eqref{eq:unitarycond} gives us a general restriction on the unitaries for which we can safely draw conclusions about the coherence properties of the system based merely on the observation of the probe.

{\it Coherent scenario}.---We can easily verify that Eq.~\eqref{eq:unitarycond} ensures that the output in Eq.~\eqref{eq:UP} remains incoherent for initially incoherent states of both system and probe.
It is now evident that whenever the output probe state is verified to be in a coherent superposition of states in $\B\s{inc;p}$,
that must be due to the initial state of the system being coherent with respect to the incoherent basis $\B\s{inc;s}$.
In other words, the unitary interaction $\Upsilon\s{ps}$ transfers the coherence of the system to the probe.
Indeed, this scenario is straightforward: 
it corresponds to a coherence generating map $\Lambda$ in Eq.~\eqref{eq:UP} that transforms an incoherent input (probe) state into a coherent output (probe) state.
We are thus mainly interested in the challenging case in which the map $\Lambda$ is incoherent.

{\it Incoherent scenario}.---Whenever the output probe state of the process in Eq.~\eqref{eq:UP} is incoherent, the naive conclusion is that no signature of the system's initial coherence survives the {\it incoherent} process $\Lambda$.
In the following, we show that this conclusion is not correct.

{\it Schr\"{o}dinger versus Heisenber pictures}.---According to the standard quantum mechanics, the probability of outcome $k$ in the measurement $\MM$ on the output of the channel $\Lambda$ for the input state $\varrho\su{in}$ is given by the Born rule, that is,
\begin{equation}\label{eq:BornS}
    p(k|\varrho\su{in},\Lambda,\MM)=\Tr(M_k\Lambda[\varrho\su{in}])=\Tr(M_k\varrho\su{out}),
\end{equation}
where the map $\Lambda$ is defined most generally through its Kraus operators as introduced in Eq.\eqref{eq:channel}.
For the special case of a unitary channel as in Eq.~\eqref{eq:unitarychannel}, this reduces to
\begin{equation}\label{eq:UBornS}
    p(k|\varrho\su{in},\Upsilon,\MM)=\Tr(M_k\Upsilon[\varrho\su{in}])=\Tr(M_k U\varrho\su{in} U^\dag).
\end{equation}
We now recall from elementary quantum mechanics that Eq.~\eqref{eq:UBornS} describes the Born rule in the Schr\"{o}dinger picture wherein the state of the system undergoes the dynamical evolution according to the unitary $\Upsilon$ and the effect $M\s{k}$ remains stationary.

We know, however, that the unitary quantum evolution can also be expressed in the Heisenberg picture by using the rule of permutation-under-the-trace as
\begin{equation}\label{eq:UBornH}
    p(k|\varrho\su{in},\Upsilon,\MM)=\Tr( U^\dag M_k U\varrho\su{in}),
\end{equation}
wherein the dynamical evolution is associated with the observable rather than the input state of the system.

Similarly to Eq.~\eqref{eq:UBornH}, the case of a general channel of Eq.~\eqref{eq:BornS} can also be recast in the Heisenberg picture as
\begin{equation}\label{eq:BornChan}
    p(k|\varrho\su{in},\Lambda,\MM)=\Tr(M_k\Lambda[\varrho\su{in}])=\Tr(\Lambda\ud[M_k]\varrho\su{in}).
\end{equation}
Here, $\Lambda\ud$ is called the {\it dual channel} to $\Lambda$.
One can easily workout the relation between $\Lambda$ and its dual $\Lambda\ud$ in the above equation (see SI-2) to find
\begin{equation}\label{eq:dualchannel}
    \Lambda\ud[M_k]:=\sum_i K_i^\dag M_k  K_i.
\end{equation}
It can be easily checked that Eq.~\eqref{eq:dualchannel} reduces to the usual passage from the Schr\"{o}dinger to the Heisenberg picture for a unitary channel.

Despite the similarities between a unitary and a generic channel in Schrodinger and Heisenberg representations, the two have a very sharp contrast.
Suppose $\Upsilon$ is an \textit{incoherent} unitary channel with respect to the incoherent basis $\B\s{inc}$.
Because $\Upsilon$ is invertible, it must transform pure states in $\B\s{inc}$ to only pure states in $\B\s{inc}$.
The latter must also hold for the inverse channel $\Upsilon^{-1}[\cdot]= U^{-1}\cdot U$.
Using the fact that $ U^{-1}= U^\dag$ we find $\Upsilon^{-1}[\cdot]= U^\dag\cdot U=\Upsilon\ud[\cdot]$.
Thus, $\Upsilon\ud[\cdot]$ must also transform pure states in $\B\s{inc}$ to only pure states in $\B\s{inc}$, i.e., the dual of an incoherent unitary channel is also an incoherent unitary channel.
This simple correspondence between the coherence properties of a unitary channel and its dual, however, does not hold for a generic channel~\cite{Chitambar2016-2}.
In other words, there are incoherent channels whose dual is not incoherent.
We now move on to show that this asymmetry can be exploited to prove the coherence of the initial state of an inaccessible system while the input and output probe are both incoherent. 
%%%%
\begin{figure}[t!]
\centering
\includegraphics[width=0.8\linewidth]{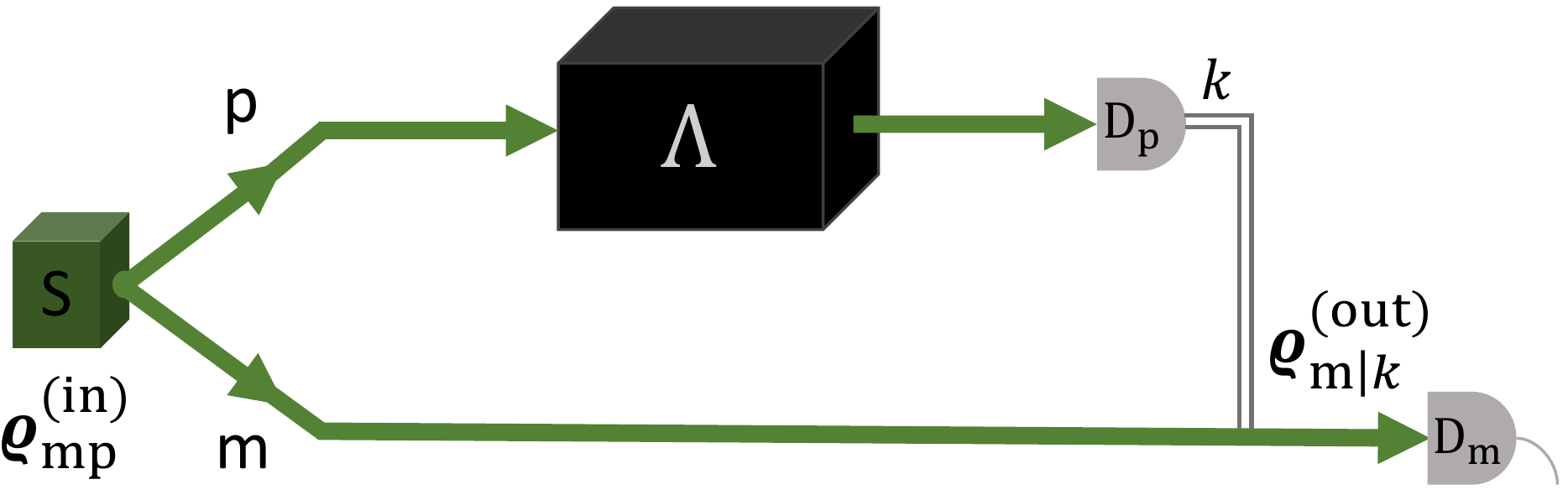}
\caption{Schematic of our coherence detection protocol.
The probe and monitor are initially prepared in the maximally entangled state $\varrho\s{mp}\su{in}$.
The probe mode then interacts with the directly inaccessible system which is part of the channel $\Lambda$.
We observe the coherence emerging in the state of the monitor conditioned on the outcomes of an incoherent measurement on the probe.}
\label{fig:scheme}
\end{figure}
%%%%
{\it The protocol}.---A twin-mode probe is initially prepared in the maximally entangled state $\varrho\s{mp}\su{in}=\sketbra{\Phi^+}{mp}$ where
\begin{equation}\label{eq:IniSProbe}
    \ket{\Phi^+}\s{mp}=n^{-1/2}\sum_{i=0}^{n-1} \ket{i}\s{m}\ket{i}\s{p}.
\end{equation}
Here,  $\ket{i}_{x}{\in}\B\s{inc;x}$ with $x{=}{\rm m},{\rm p}$, $\B\s{inc;m}{=}\B\s{inc;p}$ and $n$ is any natural number between two and infinity.
The first mode is called the monitor and the second mode is the probe to be interacted with the system.
Then, the probe mode is sent through the channel $\Lambda$ to interact with the system.
It is important to note that the quantum state of the probe mode given by the marginal density operator $\Tr\s{m}\varrho\s{mp}\su{in}=\sum_{i=0}^{n-1} \ket{i}\s{p}\bra{i}/n$ is incoherent, as required.

After the interaction between the probe and the system, an incoherent measurement on the probe $\rm p$ is carried out.
In the postprocessing stage, the outcomes of the measurement on the monitor are sorted conditioned on the outcomes of the measurement on the probe.
Any coherence within the conditional data, i.e., coherence within the \textit{conditional} output state $\varrho\sn{m}{k}\su{out}$ for the outcome $k$ of the probe, indicates that the initial state of the system $\pi\s{s}\su{in}$ was in a coherent quantum superposition; see Fig.~\ref{fig:scheme}.

In order to understand the working principles of our protocol, suppose that the channel $\Lambda\s{p}$ induced by the probe-system interaction as in Eq.~\eqref{eq:UP} is asymmetric with respect to the Schr\"{o}dinger and Heisenberg pictures, i.e., its dual $\Lambda\ud\s{p}$ is not incoherent.
We proceed step by step according to the protocol.
We have
\begin{equation}
\begin{split}
     \varrho\s{mp}\su{out} = \id\s{m}\otimes\Lambda\s{p}[\varrho\s{mp}\su{in}] & = \frac{1}{n} \sum_{i,j,l} \ket{i}\s{m}\bra{j} \otimes  K_l \ket{i}\s{p}\bra{j}  K_l^\dag\\
    & = \frac{1}{n} \sum_{i,j,l}  K_l\trs \ket{i}\s{m}\bra{j} K_l^* \otimes  \ket{i}\s{p
    }\bra{j}.
\end{split}
\end{equation}
In the last step we have used the trick $\id\otimes\Lambda[\Phi^+] = \Lambda\trs\otimes\id [\Phi^+]$ where $\Phi^+$ is the shorthand for $\ket{\Phi^+}\bra{\Phi^+}$ and $\mathsf{T}$ is the transposition operation (see SI-3 for the proof).
By postselecting on the outcome $k$ of an incoherent measurement on the probe we obtain the conditional state
\begin{equation}
\begin{split}
    \varrho\sn{m}{k}\su{out} = \sandwich{k}{\varrho\s{mp}\su{out}}{k}\s{p}
    = \frac{1}{n} \sum_{l}  K_l\trs \ket{k}\s{m}\bra{k} K_l^*.
\end{split}
\end{equation}
Finally, we examine the off-diagonal elements of the monitor's conditional state, that is,
\begin{equation}\label{eq:offdiag-condstate}
\begin{split}
    \sandwich{i}{\varrho\sn{m}{k}\su{out}}{j}\s{m} &=  \frac{1}{n} \sum_{l} \bra{i} K_l\trs \ket{k}\s{m}\bra{k} K_l^*\ket{j}\\
    &= \frac{1}{n} \sum_{l} \bra{j} K_l^\dag\ket{k}\s{m}\bra{k} K_l \ket{i},
\end{split}
\end{equation}
where we have simply transposed each matrix element and rearranged the terms.
If all the off-diagonal elements \sandwich{i}{\varrho\sn{m}{k}\su{out}}{j} of the conditional state are zero, that is, we are unable to observe any conditional coherences within the monitor, this is equivalent to stating that $ \sum_{l} \bra{j} K_l^\dag\ket{k}\s{m}\bra{k} K_l \ket{i}=0$ for all $k$ and all $i\neq j$.
This readily means that the dual channel $\Lambda\ud[\cdot]=\sum_{l}  K_l^\dag\cdot K_l$ is incoherent (see Ref.~\cite{Chitambar2016-2} or SI-4 for a proof) which contradicts our assumption.
Therefore, at least one of the conditional states $\varrho\sn{m}{k}\su{out}$ must be coherent.

As we see, the detection power of our protocol is independent of the channel being incoherent, rather it depends on the coherence properties of the dual of the channel.
Now, since (i) the probe mode is incoherent---observe that $\Tr_m \ketbra{\Phi^+}=I/n$, (ii) the channel $\Lambda$ is also incoherent, and (iii) the conditional \textit{input} monitor states $\braket{k}{\Phi^+}\braket{\Phi^+}{k}\s{p}=\sketbra{k}{m}/n$ are incoherent, we must have that the observed conditional coherence is due to the initial coherence of the system's initial state $\pi\s{s}\su{in}$.

\textit{Example.---}
Let us demonstrate our findings via a simple physical example.
Suppose the system, the probe, and the monitor are two-level systems.
We assume there are two reservoirs inducing decoherence, one of which is coupled to the system and the other couples to the probe, and investigate two regimes.

First, we consider the case where the probe-system interaction is fast enough so that system's decoherence during the interaction is negligible.
However, we assume that probe mode fully dephases due to its interaction with its reservoir before we measure it.
In this case, $\Lambda\s{p} [\varrho\s{p}\su{in}]=\Delta\s{p}\circ\Omega\s{p}[\varrho\s{p}\su{in}]$ in which $\Omega\s{p}[\varrho\s{p}\su{in}]=\Tr\s{s}\Upsilon\s{ps}[\varrho\s{p}\su{in}\otimes \pi\s{s}\su{in}]$.

Second, we analyze the case where either of the system or the probe can decohere during the interaction such that $\Lambda\s{p}[\varrho\s{p}\su{in}]=\Tr\s{s,r}\Upsilon\s{psr}[\varrho\s{p}\su{in}\otimes \pi\s{s}\su{in}\otimes \pi\s{r}\su{in}]$.
Here, the subscript $r$ stands for the reservoirs acting on the system and probe.
%%%%%%%
\begin{figure}[t!]
\centering
\includegraphics[width=0.8\columnwidth]{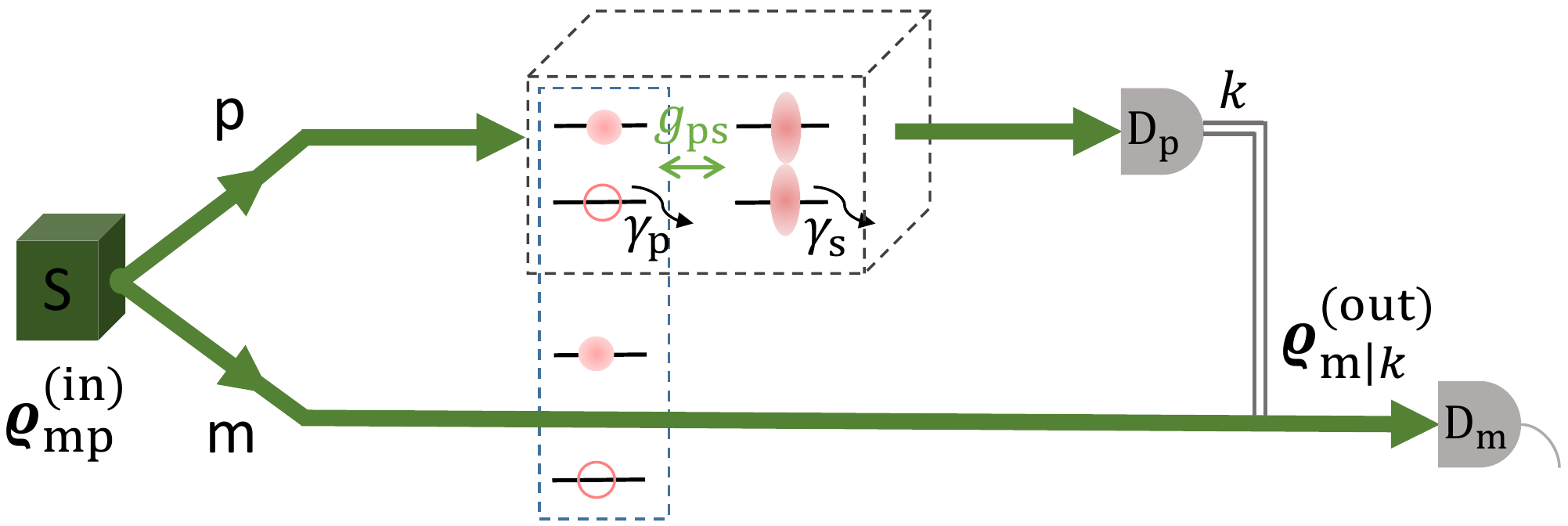}
\caption{Schematic of our example.
Two two-level systems, p and m, are initially prepared in the maximally entangled state $\varrho\s{mp}\su{in}$.
The probe mode exchanges excitation with the system, s, which is believed to be initially prepared in a superposition of its energy eigenstates, at the rate $g\s{ps}$.
While the system and probe may suffer from dephasings with rates $\gamma\s{s}$ and $\gamma\s{p}$, respectively, the monitor is isolated and decoherence-free.
Observing coherence within the state of the monitor mode conditioned on the outcomes of an incoherent measurement on the probe certifies the coherence of system's initial state.}
\label{fig_scheme_qubits}
\end{figure}
%%%%%%%%
Let $\ket{0}$ and $\ket{1}$ be the eigenstates of an arbitrary degree of freedom for the system, the probe, and the monitor.
We note that the degree of freedom may be different for each of the three subsystems.
The total Hamiltonian $(\hbar=1)$ is given by
\begin{equation}
H=H_{\rm m}+H_{\rm p}+H_{\rm s}+H_{\rm int},
\end{equation}
in which $H_{\rm j}=\omega_{\rm j}\sigma_{\rm j}^+\sigma_{\rm j}^-$ is the bare Hamiltonian of the mode ${\rm j}={\rm m, p,s}$. 
$\sigma_{\rm j}^+$ ($\sigma_{\rm j}^-$) is the raising (lowering) operator for mode ${\rm j}$ and $\omega_{\rm j}$ is the excitation energy.
The probe-system interaction is described by the Hamiltonian~\cite{scully1999book,Haroche2006book}
\begin{equation}\label{H_int}
H_{\rm int}=g_{\rm ps}(\sigma_{\rm p}^+\sigma_{\rm s}^-+\sigma_{\rm p}^-\sigma_{\rm s}^+),
\end{equation}
where $g_{\rm ps}$ is the rate of the excitation exchange between the system and the probe (see Fig.~\ref{fig_scheme_qubits}).
The unitary interaction generated by this Hamiltonian satisfies the condition in Eq.~\eqref{eq:unitarycond} which allows us to make valid statements regarding the system's initial coherence.
We further assume that there are no decay processes in the protocol that transfer the excitation of mode ${\rm j}$ to the environment, that is the collective dynamics of the probe-system preserves the total excitation number. 
This is essential if we post-select  excited states of the output probe. 
However, the probe-system dynamics is susceptible to a pure dephasing process that eliminates the phase coherence of local excitations of mode ${\rm i}={\rm p,s}$ at a rate of $\gamma_{\rm i}$. 
The total dynamics of the probe-system can be modelled by a Born-Markov master equation of the form~\cite{gardiner2004book,zhang2007dephasing,Plenio2012dephasing}
\begin{equation}\label{ME}
\frac{d\varrho}{dt}=-i[H,\varrho]+\mathcal{L}_{\rm deph}[\sigma_{z{\rm s}}]\varrho+\mathcal{L}_{\rm deph}[\sigma_{z{\rm p}}]\varrho,
\end{equation}
where $\sigma_{z{\rm i}}$ is the Pauli operator along the $z$ axis of mode ${\rm i}={\rm s},{\rm p}$.
The action of the super-operator describing the dephasing processes of mode $\rm i$ is
\begin{equation}
\mathcal{L}_{\rm deph}[\sigma_{z{\rm i}}]\varrho=\frac{\gamma_{\rm i}}{2}(\sigma_{z{\rm i}}\varrho\sigma_{z{\rm i}}-\varrho).
\end{equation}

{\it First regime}.--- 
The system is initially in an arbitrary pure state of the form $\ket{\phi(0)}\s {s}=\epsilon\ket{1}\s{s}+\sqrt{1-\epsilon^2}\ket{0}\s{s}$ where $\epsilon$ determines its degree of coherence.
We simulate the performance of our protocol in this scenario in two steps.
First, we let the probe interact with the system via the interaction Hamiltonian (\ref{H_int}).
Next, we allow the probe to fully dephase after the interaction, followed by a measurement of the coherence in the monitor conditioned on finding the probe in the excited state $\ket{1}\s{p}$. 
The details of the calculations for this simulation are provided in SI-5.
%
%%%%%
\begin{figure}[t!]
\centering
\includegraphics[width=0.9\columnwidth]{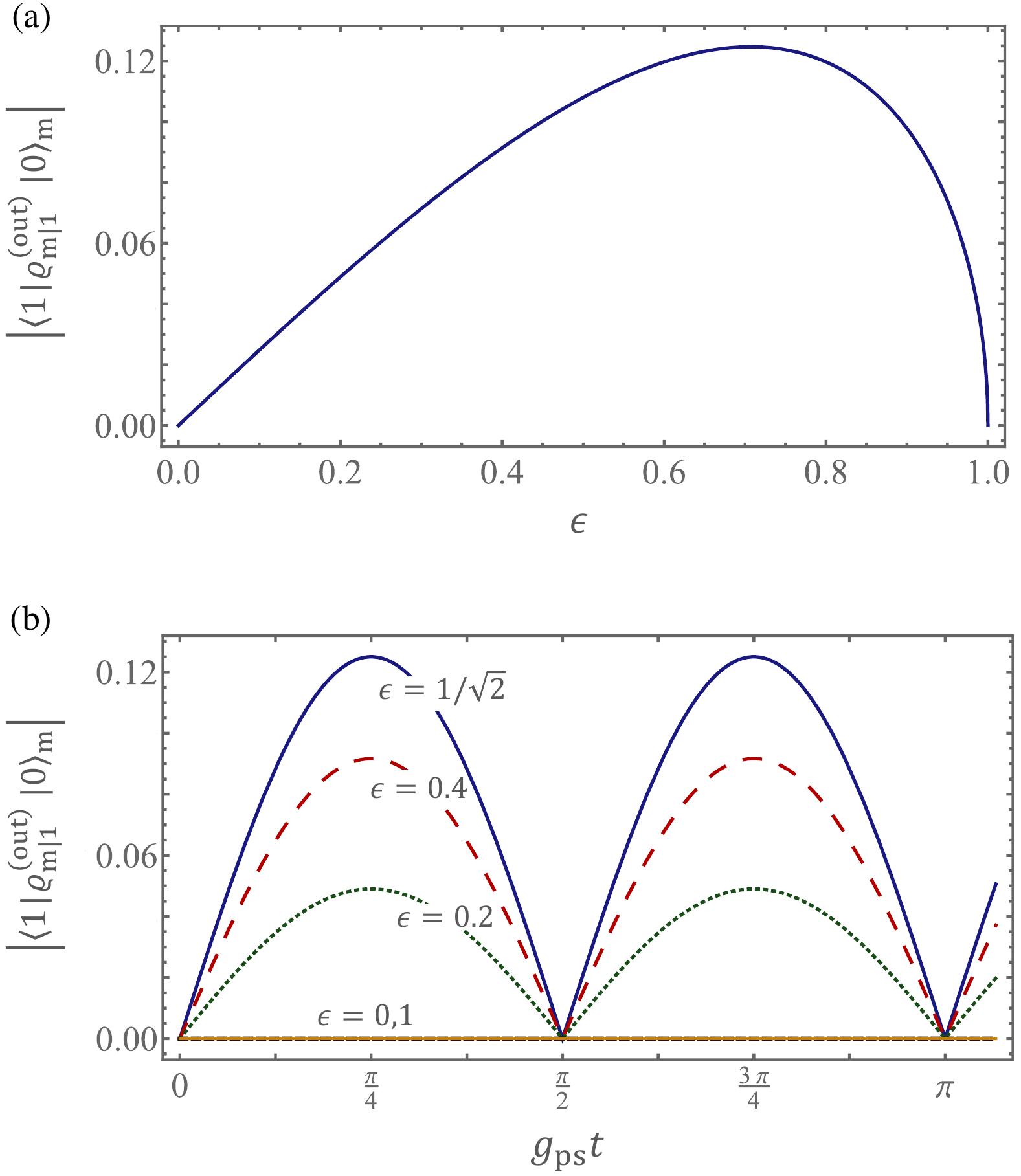}
\caption{The evolution of quantum coherence within the state of the system in the first regime. 
The magnitude of the monitor's off-diagonal density matrix element conditioned on detecting a single excitation in the probe is depicted (a)~versus the degree of the system's initial coherence, $\epsilon$, at $g_{\rm ps}t=0.75$;
(b)~versus the normalized detection time, $g_{\rm ps}t$, for different values of $\epsilon$.}
\label{fig4}
\end{figure}
%%%%%
%

Figure \ref{fig4} (a) shows the magnitude of the monitor's off-diagonal element of the density matrix, $|\sandwich{1}{\varrho\sn{m}{1}\su{out}}{0}\s{m}|$, versus system's degree of coherence $\epsilon$.
For $\epsilon=0$ and $\epsilon=1$, where the system is in an incoherent state, the monitor remains in an incoherent state as expected. 
However, for any other non-zero $\epsilon$, where the system contains initial coherence, a measurement of the off-diagonal element of the monitor results in a non-zero value with its maximum occurring at $\epsilon=1/\sqrt{2}$, i.e. for the maximally-coherent initial state of the system.
Figure~\ref{fig4}~(b) shows the magnitude of the monitor's off-diagonal element of the density matrix versus the detection time for different values of $\epsilon$.
These results clearly show that in situations where the probe is prone to dephasing before we measurement it, our protocol successfully detects the initial coherence within the system. 

The effect of probe's dephasing before the interaction with the system can be seen as a reduction in the initial quantum correlations between the probe and monitor.
It is both natural and correct that such a reduction in correlations reduces the power of the monitor to accumulate the coherence exhibited by the probe.
We provide the detail of this analysis in SI-6.
%
%%%%%%
\begin{figure}[t!]
\centering
\includegraphics[width=0.9\columnwidth]{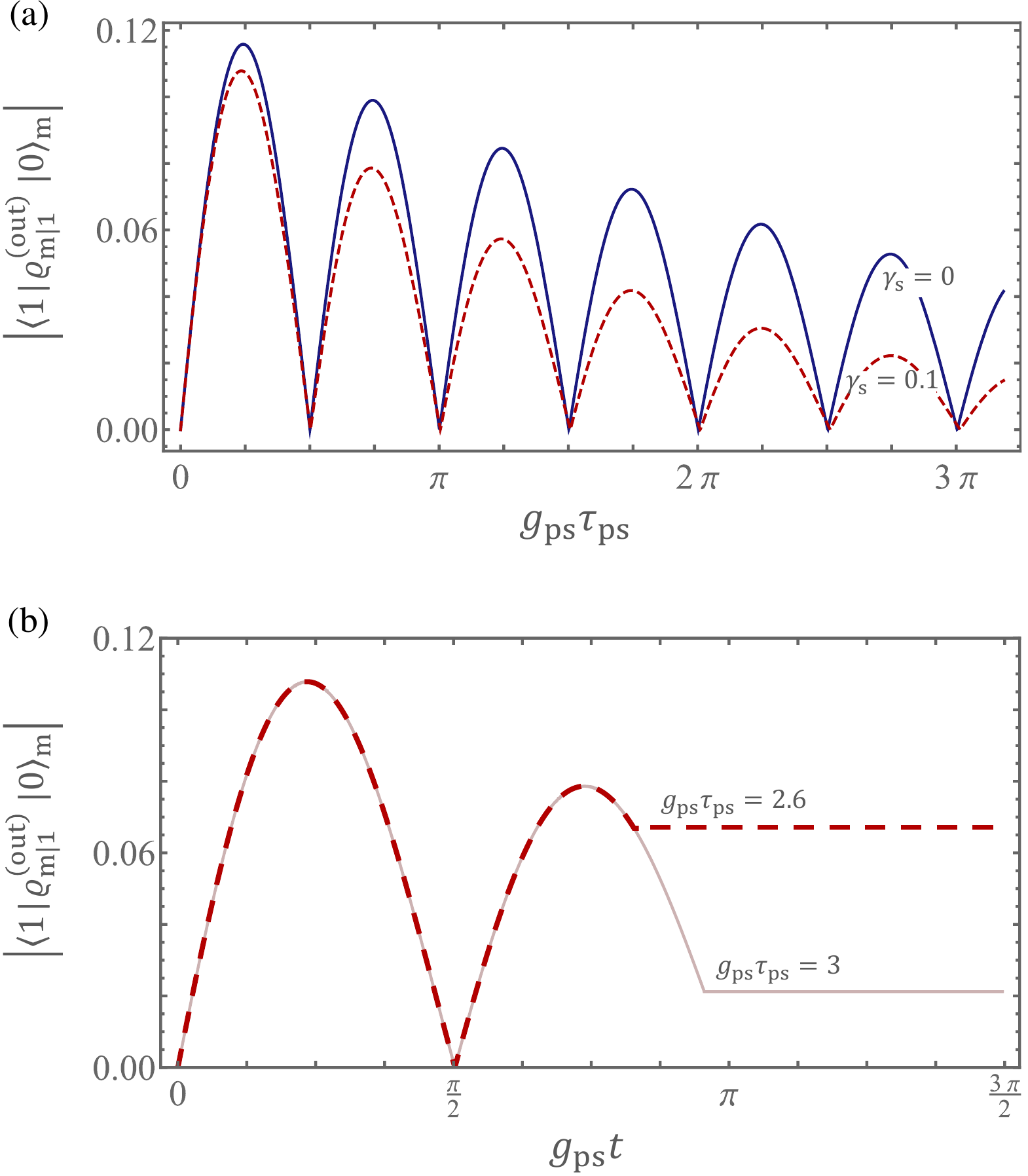}
\caption{The evolution of quantum coherence within the state of the system in the second regime.
The magnitude of the monitor's off-diagonal density matrix element conditioned on detecting a single excitation in the probe is depicted (a)~versus the normalised probe-system interaction duration, $g_{\rm ps}\tau\s{ps}$, for the cases where the probe alone decoheres during the interaction, i.e., $\gamma\s{p}=0.1$ and $\gamma\s{s}=0$ (solid blue line), and where both the probe and the system decohere while interacting, i.e., $\gamma\s{p}=0.1$ and $\gamma\s{s}=0.1$ (dashed red line);
(b)~versus the normalized evolution time of the monitor for $\gamma\s{s}=\gamma\s{p}=0.1$ and two different values of interaction duration.
In both graphs the probe continues to dephase after $\tau\s{ps}$ until the measurement is performed at an arbitrary time $t$. }
\label{fig5}
\end{figure}
%%%%%
%
{\it Second regime}.--- 
Let us now consider the more realistic regime in which the system and probe decohere during their interaction due to their coupling to the environment.
We are particularly interested in the relative time scales of the coherent and dephasing evolutions which allow for detecting the initial coherence of the system.

We consider the same initial states of $\ket{\phi(0)}\s{s}$ and $\ket{\Phi^+}\s{mp}$ for the system and monitor-probe, respectively.
We let the system and probe interact for the duration $\tau\s{ps}$ in which both system and probe are susceptible to decoherence.
The dynamical evolution of the entire system is simulated by solving the master equation~\eqref{ME} for the total state $\varrho\s{mps}(t)$. 
Then, we let the probe alone decohere further for the duration $\tau\s{meas}$ until it is being measured.
Finally, we trace over the state of the system which is inaccessible and calculate the monitor's off-diagonal element $\sandwich{1}{\varrho\sn{m}{1}\su{out}}{0}\s{m}$, i.e., conditioned on finding the probe in the excited state $\ket{1}\s{p}$ .

In Figure \ref{fig5}~(a), we depict monitor's coherence as a function of the probe-system interaction time.
It can clearly be seen that, for fast enough probe-system interactions such that $\tau\s{ps}<{\rm min}(1/\gamma\s{p},1/\gamma\s{s})$, the initial coherence of the system will be successfully observed in the monitor even if the probe fully dephases during its free evolution time $\tau\s{meas}$. 
However, if $\tau\s{ps}$ is of the order of the dephasing time of either system or probe, that is $\tau\s{ps}\approx {\rm min}(1/\gamma\s{p},1/\gamma\s{s})$, then the monitor will not retain the coherence information.
Figure~\ref{fig5}~(b) illustrates the time evolution of the induced coherence within the monitor for this case. 

We observe that decoherence occurring during the probe-system interaction causes loss of coherence information in the monitor manifested in the reduction of $|\sandwich{1}{\varrho\sn{m}{1}\su{out}}{0}\s{m}|$.
This is better understood if we think of the probe as an infinitesimally small part of the bath coupled to the system that is accessible to us.
Upon the system's interaction with its surroundings, its coherence information spreads to the entire environment.
As a result, the longer the probe-system interaction is, the smaller the share of the probe from that information will be.
Hence, for the monitor to preserve the information on the initial coherence of the system, the probe-system interaction time should be shorter than the coherence lifetime of the system.
Counter-intuitively, further decoherence of the probe does not affect the monitoring of the system's  coherence 
because, with the help of quantum correlations between the probe and monitor, a copy of this information is stored in the monitor, which is decoherence-free.

In conclusion, we have shown the counterintuitive phenomenon of detecting a quantum system's initial coherence when the input and output probe states are completely incoherent. 
We achieved this through rigorously examining the elements of coherence detection schemes from a quantum information perspective. 
Our analysis yields the necessary and sufficient conditions to enable valid claims regarding the coherence of a directly inaccessible system. 
We provided a protocol for witnessing the quantum coherence of such systems that satisfies these conditions. 
This protocol is then used in a proof-of-principle demonstration of our results, highlighting its power and limitations.
Our results confirm the necessity of fast measurements when it comes to the duration of the probe's interaction with the system.

We believe our analysis inspires novel protocols to detect coherent channels using entangled probes.
The use of a quantum entangled twin-mode probe also opens up an avenue for experimental schemes in which the probe might undergo decoherence before its information is retrieved.
In particular, we think that, in the near future, entangled light will play a significant role in probing all sorts of quantum phenomena, including quantum coherence, in complex systems.

\begin{acknowledgments}
The authors gratefully acknowledge Eric Chitambar, Eric Bittner, Martin Ringbauer, Ivan Kassal, and Eric Gauger for helpful discussions.
S.B-E. acknowledges funding from the European Union Horizon 2020 research and innovation programme under the Marie Sk\l{}odowska-Curie grant agreement No 663830.
F.S. was supported by the Royal Commission for the Exhibition of 1851 Research Fellowship. 
This project was also supported in part through the S\^{e}r SAM Project at Swansea University, an initiative funded through the Welsh Government’s S\^{e}r Cymru II Program (European Regional Development Fund).
\end{acknowledgments}

\bibliography{CoherenceBib}
\bibliographystyle{apsrev4-1new}

\end{document}

\title{Supporting Information: Incoherent witnessing of quantum coherence}

\author{Sahar Basiri-Esfahani}
 \email{Electronic address: sahar.basiriesfahani@swansea.ac.uk}
 \affiliation{Department of Physics, Swansea University, Singleton Park, Swansea SA2 8PP, United Kingdom}

 \author{Farid Shahandeh}
 \email{Electronic address: shahandeh.f@gmail.com}
 \affiliation{Department of Physics, Swansea University, Singleton Park, Swansea SA2 8PP, United Kingdom}

\maketitle

\section*{SI-1: Kraus decomposition of quantum channels  }

A generic channel $\Lambda$ transforming an input state to the output can be written as~\cite{NielsenBook}
\begin{equation}\label{eq:channel}
    \varrho\su{out}=\Lambda [\varrho\su{in}]=\sum_i K_i\varrho\su{in} K_i^\dag.
\end{equation}
Here, the operators $ K_i$ are called Kraus operators of the channel and satisfy $\sum_i  K_i^\dag  K_i{=}\Ident$.
To see that such a transformation maps density operators to density operators, we first show that the output operator is positive.
For an \textit{arbitrary} quantum state $\ket{\psi}$ we have
\begin{equation}\label{eq:posit}
    \bra{\psi}\varrho\su{out}\ket{\psi}=\sum_i \bra{\psi}K_i\varrho\su{in} K_i^\dag\ket{\psi}.
\end{equation}
It is clear that $\bra{\phi}:=\bra{\psi}K_i$ is the Hermitian conjugate of $\ket{\phi}:=K_i^\dag\ket{\psi}$.
Since $\varrho\su{in}$ is a positive operator, it follows that $\bra{\phi}\varrho\su{in}\ket{\phi}\geq 0$.
Thus, $\bra{\psi}\varrho\su{out}\ket{\psi}\geq 0$ which proves our claim.

Now we show that channel $\Lambda$ preserves the normality of the density operators:
\begin{equation}\label{eq:channel}
    \Tr\varrho\su{out}=\sum_i \Tr (K_i\varrho\su{in} K_i^\dag)=\sum_i \Tr (K_i^\dag K_i\varrho\su{in}) =  \Tr (\sum_i K_i^\dag K_i\varrho\su{in}) = \Tr (\Ident\varrho\su{in}) =1.
\end{equation}
Hence, the output remains a normalized positive operator, i.e., a valid density operator.

%%%%%%%%%%%%%%%%%%%%%%%%%%%%%%%%%%%%%%%%%%%%%%%%%%%%%%%%%%%%%%%%%%%%%%%%%%%%%%%%%%%%%%%%%%%%%%%%%%%%%%%%%%%%%%
\section*{SI-2: Relation between a quantum channel and its dual channel}

Here, we show that the dual of a channel $\Lambda$ with Kraus decomposition $\Lambda [\cdot] = \sum_l K_i [\cdot] K_i^\dag$ is given by $\Lambda\su{d} [\cdot] = \sum_l K_i^\dag [\cdot] K_i$.
Consider the following sequence of operations:
\begin{equation}
\begin{split}
    \Tr(\hat{M}_k\Lambda[\varrho\su{in}])&=\Tr\left(\hat{M}_k\sum_i \hat{K}_i\varrho\su{in}\hat{K}_i^\dag\right)\\
    & = \sum_i\Tr\left(\hat{M}_k \hat{K}_i\varrho\su{in}\hat{K}_i^\dag\right)\\
    & = \sum_i\Tr\left(\hat{K}_i^\dag \hat{M}_k \hat{K}_i \varrho\su{in}\right)\\
    & = \Tr\left(\sum_i\hat{K}_i^\dag \hat{M}_k \hat{K}_i \varrho\su{in}\right)\\
    & = \Tr(\Lambda\ud[\hat{M}_k]\varrho\su{in}).
\end{split}
\end{equation}
Since this is true for any input state, it follows from the last equality that
\begin{equation}
    \Lambda\su{d} [\cdot] = \sum_l K_i^\dag [\cdot] K_i.
\end{equation}

%%%%%%%%%%%%%%%%%%%%%%%%%%%%%%%%%%%%%%%%%%%%%%%%%%%%%%%%%%%%%%%%%%%%%%%%%%%%%%%%%%%%%%%%%%%%%%%%%%%%%%%%%%%%%%%%%%%
\section*{SI-3: Proof of $\id\otimes\Lambda[\Phi^+] = \Lambda\trs\otimes\id [\Phi^+]$ }

Given an operator $\hat{K}$ and a basis $\{\ket{i}\}$,
\begin{equation}
\begin{split}
\hat{I}\otimes\hat{K}\ket{i}\s{A}\ket{i}\s{B} = \ket{i}\s{A}\sum_j \ket{j}\s{B}\sandwich{j}{\hat{K}}{i}\s{B}.
\end{split}
\end{equation}
Then, we use the facts that (i) the two Hilbert spaces of A and B are assumed to be isomorphic, $\HH\s{A}\cong\HH\s{B}$, and (ii) $\s{B}\sandwich{j}{\hat{K}}{i}\s{B}$ is a c-number and thus we can safely replace it with $\s{A}\sandwich{j}{\hat{K}}{i}\s{A}$, to write
\begin{equation}\label{eq:AtoBParts}
\begin{split}
\hat{I}\otimes\hat{K}\ket{i}\s{A}\ket{i}\s{B} &= \sum_j \ket{i}\s{A}\sandwich{j}{\hat{K}}{i}\s{A} \ket{j}\s{B}\\
& = \sum_j \ket{i}\s{A}\sandwich{i}{\hat{K}\trs}{j}\s{A} \ket{j}\s{B}\\
& = n^{\frac{1}{2}} \ket{i}\s{A}\bra{i}\left(\hat{K}\trs\otimes\hat{I}\right)\ket{\Phi^{+}}\s{AB},
\end{split}
\end{equation}
where $\ket{\Phi^{+}}\s{AB}$ possesses the computational basis as its Schmidt vectors.
It is now easy to sum over $i$ on both sides of Eq.~\eqref{eq:AtoBParts} to obtain the desired result,
\begin{equation}
\begin{split}
& \hat{I}\otimes\hat{K}\sum_i\ket{i}\s{A}\ket{i}\s{B} = n^{\frac{1}{2}} \sum_i \ket{i}\s{A}\bra{i}\left(\hat{K}\trs\otimes\hat{I}\right)\ket{\Phi^{+}}\s{AB} \\
\Rightarrow & \hat{I}\otimes\hat{K} \ket{\Phi^{+}}\s{AB}  = \hat{K}^T\otimes\hat{I}\ket{\Phi^{+}}\s{AB}.
\end{split}
\end{equation}
%
%%%%%%%%%%%%%%%%%%%%%%%%%%%%%%%%%%%%%%%%%%%%%%%%%%%%%%%%%%%%%%%%%%%%%%%%%%%%%%%%%%%%%%%%%%%%%%%%%%%%%%%%%%%%%%%%%%%
\section*{SI-4: necessary and sufficient condition for incoherent channels}

For a quantum channel $\Lambda$ with Kraus decomposition $\Lambda [\cdot] = \sum_l K_l [\cdot] K_l^\dag$, the necessary and sufficient condition to be incoherent is given by~\cite{Chitambar2016-2}
\begin{equation}\label{eq:MIOiff}
\begin{split}
\sum_{l}\bra{j} K_l\sketbra{i}{p} K_l^\dag\ket{k}=0,
\end{split}
\end{equation}
for all $i$ and $j\neq k$.

To show this, we use a representation of quantum channel known as the Choi-jamio\l{}kowski isomorphism~\cite{Jamiolkowski1972,Choi1975,Jiang2013}.
For our purpose, it is enough to recall that this isomorphism is equivalent to applying the channel to the maximally entangled state $\ket{\Phi^+}\bra{\Phi^+}$, represented as $\Phi^+$ for short.
Now, using the definition of an incoherent channel given in the main text, that is $\Lambda\circ\Delta[\cdot] = \Delta\circ\Lambda\circ\Delta[\cdot]$, we find the Choi-jamio\l{}kowski representation of both sides as
\begin{equation}
\begin{split}
\Lambda\circ\Delta[\Phi^+] &= \sum_{i,l}\sketbra{i}{m}\otimes  K_l\sketbra{i}{p} K_l^\dag
\end{split}
\end{equation}
and
\begin{equation}
\begin{split}
\Delta\circ\Lambda\circ\Delta[\Phi^+] &= \sum_{i,j,l}\sketbra{i}{m}\otimes \sketbra{j}{p} \bra{j} K_l\sketbra{i}{p} K_l^\dag\ket{j}.
\end{split}
\end{equation}
Note that, the fully dephasing channel and $\Lambda$ only act on the probe mode in $\Phi^+$.
Equating the two relations, it follows that
\begin{equation}\label{eq:MIOiff}
\begin{split}
\sum_{l}\bra{j} K_l\sketbra{i}{p} K_l^\dag\ket{k}=0,
\end{split}
\end{equation}
as claimed.
%%%%%%%%%%%%%%%%%%%%%%%%%%%%%%%%%%%%%%%%%%%%%%%%%%%%%%%%%%%%%%%%%%%%%%%%%%%%%%%%%%%%%%%%%%%%%%%%%%%%%%%%%%%%%%%%%%%
\section*{SI-5: Detecting coherent state of a two-level system: first regime}

Here, we calculate the conditional state of the monitor conditioned on finding the probe in the excited state $\ket{1}\s{p}$. To do this, we first calculate the total state of the probe-system after the interaction. This will be $\varrho\s{mps}(t)=\ket{\Phi(t)}\s{mps}\bra{\Phi(t)}$, where
\begin{equation}
\begin{split}
\ket{\Phi(t)}\s{mps}&=I\s{m}\otimes U\s{ps}(t)\ket{\Phi^+}\s{mp}\ket{\phi(0)}_ {\rm s}\\
& = \frac{1}{\sqrt{2}}\sum_{i=0}^{1} \ket{i}\s{m}\ket{\chi_i}\s{ps},
\end{split}
\end{equation}
in which $U\s{ps}(t)=e^{-iH_{\rm int}t}$ is the evolution operator in the interaction picture and
\begin{equation}
\begin{split}
\ket{\chi_i}\s{ps}&=\sum_{j=0}^{1}\ket{j}\s{p}\ket{\alpha_{ij}}\s{s},\\
\ket{\alpha\s{00}}\s{s}&=\sqrt{1-\varepsilon^2}\ket{0}\s{s}+\varepsilon\cos(g\s{ps}t)\ket{1}\s{s},\\
\ket{\alpha\s{01}}\s{s}&=-i\varepsilon\sin(g\s{ps}t)\ket{0}\s{s},\\
\ket{\alpha\s{01}}\s{s}&=-i\sqrt{1-\varepsilon^2}\sin(g\s{ps}t)\ket{1}\s{s},\\
\ket{\alpha\s{11}}\s{s}&=\sqrt{1-\varepsilon^2}\cos(g\s{ps}t)\ket{0}\s{s}+\varepsilon\ket{1}\s{s}.
\end{split}
\end{equation}
Hence, the state of the probe is $\varrho\s{mp}(t)={\rm Tr_s}(\varrho\s{mps}(t))$. Now, the action of a fully dephasing operation on the probe p-mode results in the output state
\begin{equation}
\begin{split}
\varrho\s{mp}\su{out} &=\Delta[\varrho\s{mp}(t)]\\
&=\frac{1}{2}\sum_{j=0}^{1}\ket{j}\s{p}\bra{j}\otimes\sum_{i,l=0}^{1}\langle\alpha_{lj}\vert\alpha_{ij}\rangle\ket{i}\s{m}\bra{l}.
\end{split}
\end{equation}
Next, we postselect on outcome $k$ ($k=0$ or $1$) of a projective measurement on the probe p-mode. The conditional state of the probe m-mode becomes
\begin{equation}
\begin{split}
    \varrho\sn{m}{k}\su{out} &= \sandwich{k}{\varrho\s{mp}\su{out}}{k}\s{p}\\
    & = \frac{1}{2}\sum_{i,l=0}^{1}\langle\alpha_{lk}\vert\alpha_{ik}\rangle\ket{i}\s{m}\bra{l}.
\end{split}
\end{equation}
Finally, we measure the off diagonal elements of the conditional state of the monitor mode as
\begin{equation}
    \sandwich{i}{\varrho\sn{m}{k}\su{out}}{l} =  \frac{1}{2} \langle\alpha_{lk}\vert\alpha_{ik}\rangle.
\end{equation}
Figure 4(b) in the main text shows the plot of the time evolution of the absolute value of this off-diagonal element for $i=1$ and $l=0$.
%%%%%%%%%%%%%%%%%%%%%%%%%%%%%%%%%%%%%%%%%%%%%%%%%%%%%%%%%%%%%%%%%%%%%%%%%%%%%%%%%%%%%%%%%%%%%%%%%%%%%%%%%%%%%%%%%%%
\section*{SI-6: Probe's decoherence before the interaction}
%%%%%%%%
\begin{figure}[t!]
\centering
\includegraphics[width=0.5\columnwidth]{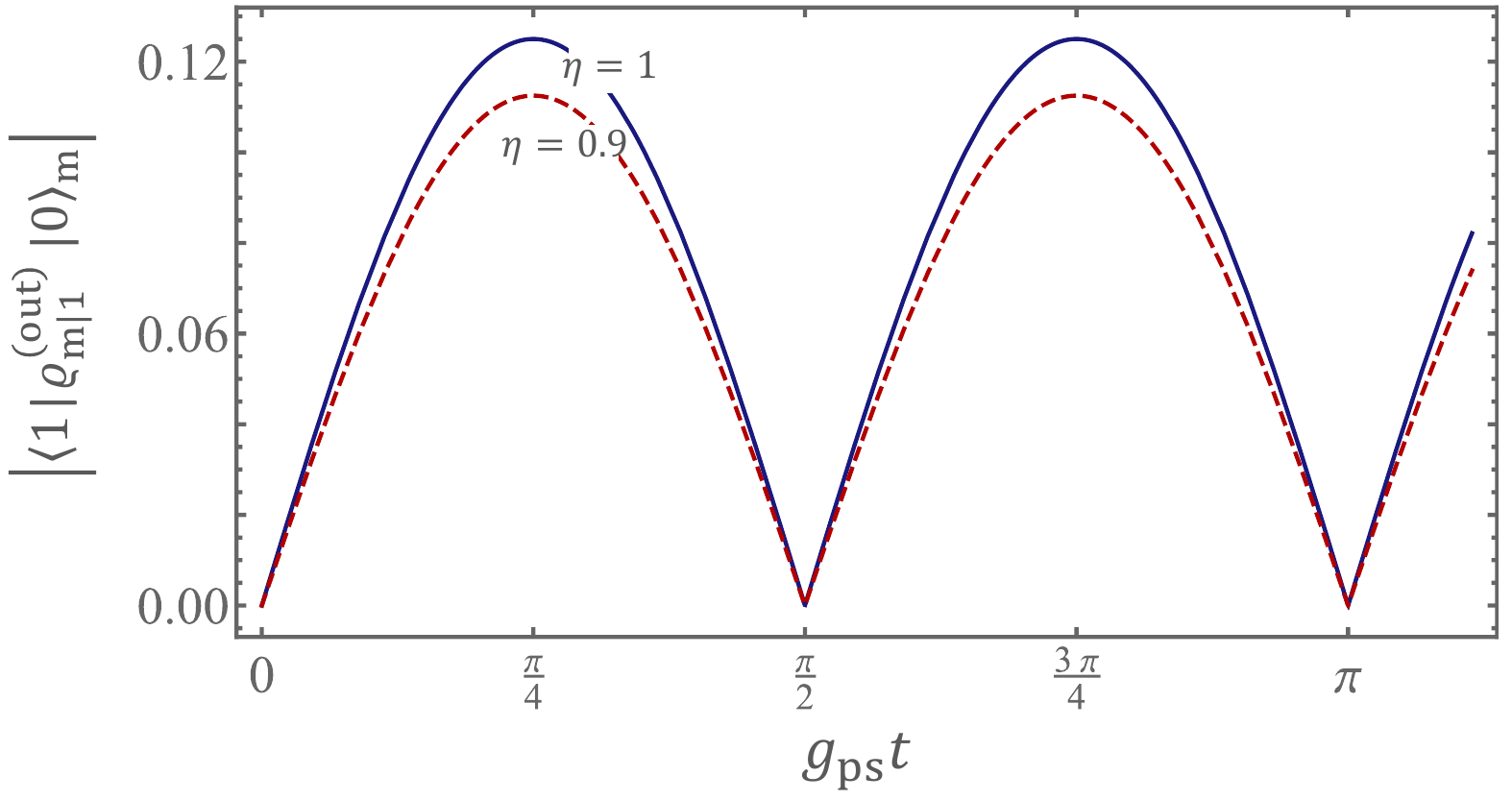}
\caption{Detected coherence for a probe which is partially dephased before its interaction with the system.
The magnitude of monitor's off-diagonal element of the density matrix conditioned on detecting a single excitation in the probe is plotted versus the normalized detection time, $g\s{ps}t$, for $\epsilon=1/\sqrt{2}$ and different values of $\eta$. }
\label{fig-SI}
\end{figure}
%%%%%%%%
We mentioned within the main text that the effect of probe being dephased before the interaction with the system can be seen as a reduction in the initial quantum correlations between the probe and monitor.
To see this, let us introduce the partial dephasing channel defined as
\begin{equation}\label{eq:partDeph}
    \Delta_\eta[\varrho]:=\eta \, I[\varrho] + (1-\eta) \, \Delta[\varrho],
\end{equation}
in which $0\leq \eta \leq 1$ and $\Delta[\varrho]= \sum_i \sandwich{i}{\varrho}{i} \ketbra{i}=\sum_i \varrho_i \ketbra{i}$ is the fully dephasing channel.
Equation~\eqref{eq:partDeph} has a simple physical meaning: it is equivalent to applying no dephasing to the state with probability $\eta$ and fully dephasing it with probability of $1-\eta$.

A two-mode probe state that undergoes decoherence before interacting with the system can be thought of as going through the partial dephasing channel $\Delta_\eta$.
We thus find for the input probe-monitor state that
\begin{equation}
\Delta_\eta[\Phi^+\s{mp}]=\frac{1}{2}\sum_{i=0}^{1}\ket{i}\s{m}\bra{i}\otimes\ket{i}\s{p}\bra{i}+\frac{\eta}{2}\sum_{i,j=0}^{1}\ket{i}\s{m}\bra{j}\otimes\ket{i}\s{p}\bra{j}.
\end{equation}
It is evident that $\eta=1$ represents no dephasing and the maximum correlations between the probe and monitor, while $\eta=0$ represents a fully dephased initial state with no correlations between the two modes. 
It is both natural and correct that such a reduction in correlations reduces the power of the monitor to accumulate the coherence exhibited by the probe.
Having an initial partial probe dephasing (a non-unity $\eta$) can be seen as a reduction in the maximum value of the magnitude of the monitor's off-diagonal element of the density matrix as shown in Fig.~\ref{fig-SI}.

\bibliography{CoherenceBib}
\bibliographystyle{apsrev4-1new}